# Combinatorial Compression in Nucleobase Self-Assembly Governed by Seeding Kinetics

Yael Kapon[a], Rotem Edri[a], and Moran Frenkel-Pinter[a,b,c*]

[a] Institute of Chemistry, The Hebrew University of Jerusalem, Jerusalem, Israel

[b] The Center for Nanoscience and Nanotechnology, The Hebrew University of Jerusalem, Jerusalem, Israel 9190401

[c] Casali Center of Applied Chemistry, The Hebrew University of Jerusalem, Jerusalem 9190401, Israel

*Corresponding author:

Dr. Moran Frenkel-Pinter
The Center for Nanoscience and Nanotechnology
Institute of Chemistry
The Hebrew University of Jerusalem
Edmond J. Safra Campus
Jerusalem 9190401, Israel
Ph: (+972)-2-6584171
*moran.fp@mail.huji.ac.il*

**Keywords:** competitive crystallization; nucleobase self-assembly; combinatorial compression; prebiotic chemistry; seeded crystallization

## Abstract

Crystallization from chemically complex mixtures is governed not only by thermodynamic stability but also by kinetic competition among structurally related components during nucleation and growth. When multiple compatible molecules compete for incorporation into early nuclei, the number of possible local recognition motifs can expand, delaying access to productive crystalline pathways. Here, we investigate this effect using supersaturated aqueous mixtures of the canonical nucleobases adenine (A), thymine (T), and uracil (U) as a minimal model for competitive supramolecular crystallization. Time-resolved turbidity measurements showed that ternary A+T+U mixtures crystallize more slowly and less efficiently than selected binary systems, despite containing individually crystallizable components. This inhibition is consistent with competition-induced kinetic frustration, where structurally similar hydrogen-bonding partners compete during early nucleation, delaying formation of a productive crystalline nucleus. We then show that preformed A+T crystalline seeds rescue ordered growth from the ternary mixture. Seeding sharply reduces the lag time, increases the apparent nucleation/onset rate, and redirects the solid product toward the A+T crystalline pathway, as confirmed by HPLC. Thus, the seed acts not only as a nucleation accelerator but as a selector of molecular composition and assembly trajectory within a multicomponent solution. We describe this process as kinetic compression: the reduction of accessible supramolecular assembly pathways through amplifications of selected productive growth pathways. These findings establish seeded crystallization as a mechanism for selecting compositional pathways in competitive molecular mixtures and suggest a physical route by which ordered molecular subsets can emerge from chemically complex environments.

## Introduction

Crystallization and molecular self-assembly organize small molecules into ordered functional nanostructures across crystal engineering,[1] supramolecular organic materials design,[2,3] and pharmaceutical crystal development.[4] Yet, most mechanistic studies are performed under simplified conditions that isolate a single target pathway. In natural chemical environments, crystallization often proceeds from multicomponent mixtures, where several assembly and nucleation pathways may be accessible simultaneously,[2,5] and rarely proceeds from a single purified solute. Additional molecules can either inhibit nucleation,[6–10] change growth kinetics,[11,12] or redirect the system toward alternative solid forms such as polymorphs[13] or co-crystals.[4,14,15] In self-assembled systems, targeting the nucleation step has been successful in kinetically directing pathway selection. The addition of a preselected seed can select specific polymorphs[16] and promote assembly.[17,18] Thus, multicomponent crystallization can be viewed not only as a challenge for materials control but also as a route for chemical-space compression: from a wide chemical space with many possible molecular interactions toward a smaller set of kinetically selected ordered molecules.

Hydrogen-bonded assemblies provide a useful subcase of competitive crystallization, where directional molecular recognition can select between closely related ordered networks.[8,19,20]

Within this class, nucleobases offer a minimal and chemically interpretable platform because they are small, biologically central heteroaromatic molecules whose hydrogen-bonding faces support both canonical and noncanonical recognition modes.[21] Ordered base-pairing and related heteromolecular recognition have been demonstrated in nucleobase and nucleobase-inspired systems, including adenine-thymine nanopatterns,[22] thymine cocrystals,[23,24] and melamine–cyanuric acid cocrystals.[25,26] Competitive crystallization has been studied in nucleobase-derived systems, particularly for modified nucleobases[27] and nucleotides[28,29] in which orthogonal complementary pairs such as adenine: thymine and guanine: cytosine were shown to selectively crystallize from the same mixture. However, this still raises a distinct question: how does crystallization proceed when several components compete within the same recognition motif, such that multiple closely related ordered states are accessible from a shared molecular pool?

Here, we ask whether seeding can overcome kinetic frustration in a competitive aqueous nucleobase mixture and thereby promote compositional selection during crystallization. We use the canonical nucleobases adenine (A), thymine (T), and uracil (U) as a simplified model system for competitive self-assembly. A can recognize and assemble with either T[30] or U,[31] creating two competing binary assembly pathways within the same molecular pool (Figure 1a,b). This system allows us to examine whether preformed crystalline seeds can bias assembly in an aqueous supersaturated environment toward one selected co-assembled product.

In the A+T binary system, repetitive A-T recognition units can form an ordered nucleus and grow into a co-assembled crystal. When U is added, the number of possible local monomer arrangements within the nucleus is expanded, creating multiple competing nucleation-level configurations within the supersaturated energy landscape (Figure 1c). We describe this expansion of accessible local arrangements as a form of combinatorial explosion at the nucleation level. This term is commonly used in complex chemical systems, where increasing the number of reactive or associating components can generate a rapidly expanding space of possible products or configurations.[32–34] Several mechanisms have been shown to counter such expansion, including environmental selection in wet–dry chemical networks,[34,35] recursive selection in mineral environments,[33] and template-based amplification in dynamic combinatorial libraries.[36]

In this study, we propose that seeded non-covalent self-assembly can act as an additional route to compression of the accessible assembly landscape. Unlike polymerization, where increasing monomer diversity can increase the number of possible products, crystallization requires propagation of a compatible ordered nucleus. Increased nucleation-level diversity can therefore inhibit growth by expanding the number of local arrangements that compete with, but do not support, ordered propagation (Figure 1c). We then test whether preformed A+T seeds can promote a selected assembly pathway from the A+T+U mixture. In this framework, seeding does not simply accelerate crystallization; it acts as a route for kinetic compression, narrowing the accessible assembly trajectories on the experimental timescale by providing a preorganized A+T interface that promotes A+T-rich growth from a complex solution (Figure 1d,e). Overall, our results show that a preformed crystalline phase can select molecular composition and assembly pathway from a competitive canonical nucleobase mixture in water, thereby compressing a multicomponent supramolecular landscape into one amplified crystalline trajectory.

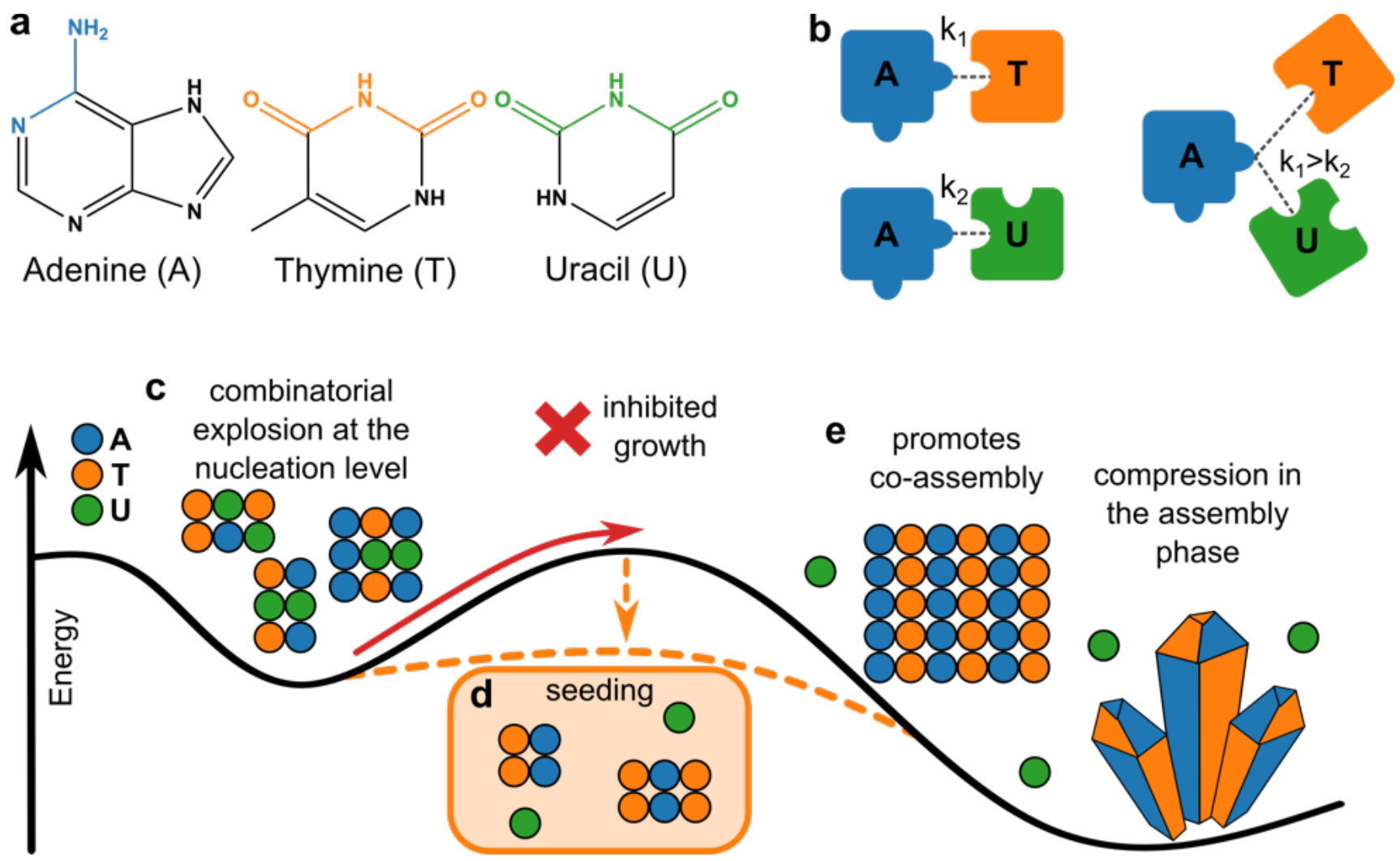

**Figure 1. Competitive nucleobase self-assembly as a route from combinatorial explosion to kinetic compression.** **(a)** We examine the competitive self-assembly of canonical nucleobases: Adenine (A), Thymine (T), and Uracil (U). These nucleobases contain Watson–Crick recognition motifs, shown in blue, orange, and green, respectively, which can compete for assembly within the same chemical space. **(b)** In binary mixtures, A can recognize and assemble with either T or U, each interaction following a distinct kinetic pathway. In complex mixtures, both competing partners are present and can participate in assembly or co-assembly. **(c)** A ternary A+T+U system expands the number of possible monomer arrangements within a nucleus, creating a combinatorial explosion at the nucleation level. This can delay access to productive ordered growth. Counterintuitively, unlike polymerization, where increased sequence diversity can generate many products, this nucleation-level complexity causes inhibited growth. **(d)** Seeding the ternary mixture with pre-formed A+T crystal seeds lowers the kinetic barrier by **(e)** promoting ordered co-assembly and compressing the accessible assembly space, directing the system toward the A+T co-crystal despite the presence of U.

## Results

A, T, and U were dissolved either as single components or as mixtures at a nucleobase concentration of 45 mM for each component, unless otherwise stated. Samples were heated to 95 °C for 3 h to ensure full dissolution, followed by cooling to room temperature. After cooling, samples were either transferred to a plate reader for turbidity measurements or incubated at room temperature for 48 h and drop-cast onto Si wafers for scanning electron microscopy (SEM) imaging. SEM was used to compare the final morphologies formed by the single-component, binary, and ternary systems. The single nucleobases each formed crystals with distinct morphologies after 48 h: A formed log-like crystals, U formed elongated crystals, and T formed flake-like crystals (Figure 2a and Figure S1). In contrast, the binary A+T mixture assembled into elongated crystalline fibers, indicating that co-assembly generates a morphology distinct from that of the individual components (Figure 2b and Figure S1). When the third component was introduced, the A+T+U mixture did not form the same ordered fibers within 48 h but instead produced a largely amorphous aggregate morphology (Figure 2c). High-resolution SEM revealed that these disordered regions nevertheless contained thin crystalline fibers, suggesting that fiber formation was not eliminated but kinetically delayed (Figure S2). Indeed, after incubation for over one week, some elongated crystalline fibers emerged from the ternary mixture, similar to those observed in the A+T binary system, along with persistent amorphous areas (Figure 2c).

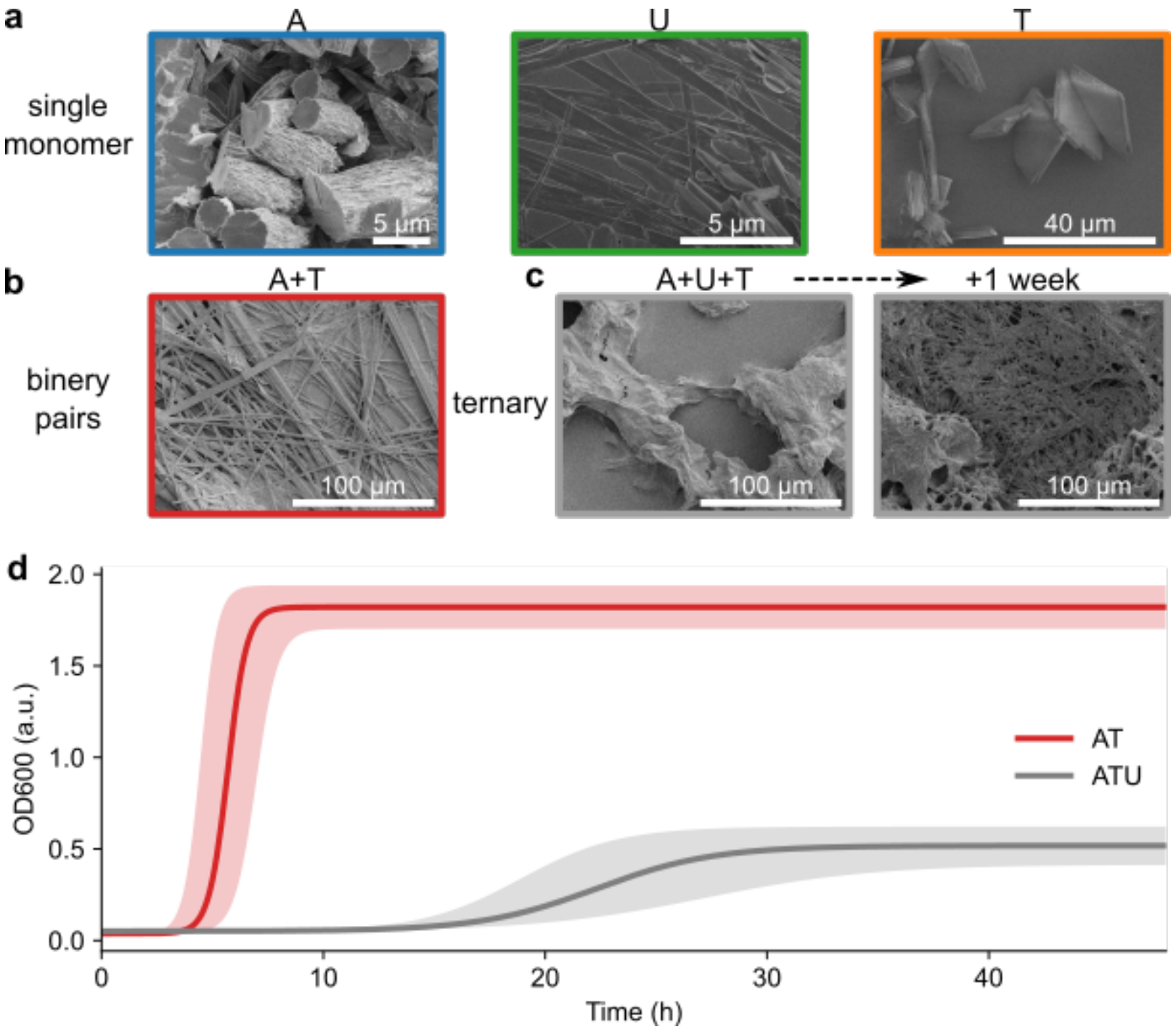

**Figure 2. Increasing mixture complexity reshapes nucleobase assembly kinetics and morphology.** (**a**) Scanning electron microscopy (SEM) images of single-component nucleobase crystals after 48 h incubation. Adenine (A) forms log-like crystals, uracil (U) forms elongated crystals, and thymine (T) forms flake-like crystals. (**b**) SEM image of the binary A+T system after 48 h incubation, showing elongated crystalline fibers with a morphology distinct from that of the individual components. (**c**) SEM images of the ternary A+T+U system after 48 h and after over one week of incubation. After 48 h, A+T+U forms largely amorphous aggregates, while after prolonged incubation, elongated crystalline fibers emerge, similar to those observed in the A+T binary system. (**d**) Turbidity kinetic traces of A+T (red) and A+T+U (gray) measured over 48 h. Curves show averaged logistic fits of replicate measurements, with shading representing the standard error of the mean. Compared with A+T, the A+T+U mixture shows inhibited assembly, with a longer lag time, lower maximum slope, and reduced saturation.

Turbidity measurements were used to monitor assembly kinetics at 600 nm. After heating and cooling, the samples were transferred to a 96-well plate, and turbidity was measured over 48 h at 30C. The resulting kinetic traces were fitted using a logistic function as a phenomenological model to compare the onset, rate, and final turbidity across samples:

$$y = b + \frac{A}{1 + e^{-k(t-t_0)}}$$

The following kinetic parameters were extracted[37,38] for each replicate well: saturation ($S = b + A$); max slope ($s_{max} = \frac{Ak}{4}$); and lag time ($t_{lag} = t_0 - \frac{2}{k}$). The lag time reflects the time required for the onset of measurable assembly (induction time + growth time to measurable turbidity), the maximum slope reflects the fastest growth rate during assembly, and the saturation reflects the final turbidity reached by the sample. Together, these parameters describe the onset, rate, and final amount of AT-like fibrillar assembly. Because turbidity reports optical scattering from suspended solids rather than molecular nucleation directly, these parameters are interpreted as comparative apparent kinetic descriptors.

For each condition, individual replicate traces were fitted separately. The extracted fitting parameters were then averaged across replicates, and the uncertainty is reported as the standard error of the mean. For visualization of the kinetic traces, the mean fitted curve is shown with the standard error of the mean represented as the shaded region. The mean fitted curves of A+T (red) and A+T+U (gray) are presented in urure 2d. The full set of individual kinetic traces, replicate fits, and averaged traces is presented in the Supporting Information (Figures S4-6).

Compared with A+T, the A+T+U mixture showed inhibited assembly of AT-like structures across all extracted kinetic parameters: a longer lag time (the lag time increased from 5±1 h to 17±3 h), lower maximum slope (decreased from 1.0±0.2 $\frac{1}{h}$ to 0.04±0.02 $\frac{1}{h}$), and reduced saturation (from 1.8±0.1 a.u. to 0.5±0.1 a.u.). These results are consistent with the largely disordered aggregates observed by SEM for the A+T+U system.

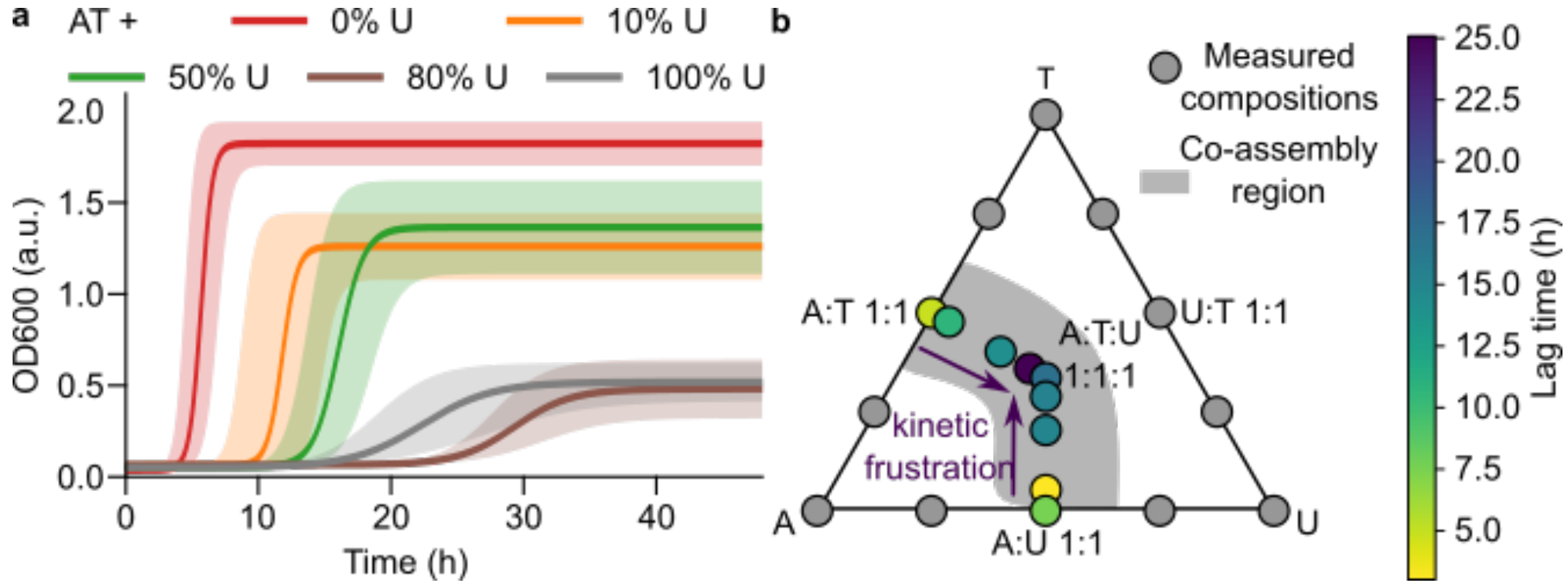

**Figure 3. A combinatorial explosion induces kinetic frustration in ternary nucleobase assembly.** (**a**) Turbidity kinetic traces of the A+T system with increasing U content, measured over 48 h. Curves show the averaged sigmoid fits of replicate measurements, with shading representing the standard error of the mean. Increasing U content progressively slows assembly kinetics. (**b**) Ternary composition map of A/T/U assembly kinetics. Color indicates the measured lag time at each composition; gray points indicate the morphologies of the measured compositions; and the shaded region marks compositions that form distinct co-assembled morphologies. Lag time increases toward the interior of the ternary composition space, indicating that combinatorial explosion within the nucleation landscape frustrates crystallization and delays assembly.

We next aimed to map how ternary composition affects assembly kinetics across the A/T/U composition space. To do so, kinetic measurements were performed for both A+T mixtures with increasing U content (Figure 3a) and A+U mixtures with increasing T content (Figures S9-S10). For the A+T series, U was added at 0%, 10%, 50%, 80%, and 100% relative to A/T (corresponding to 4.5, 22.5, 36, and 45 mM, respectively), while keeping the A and T concentrations fixed at 45 mM each. The averaged turbidity traces showed a general slowing and suppression of assembly with increasing U content (Figure 3a). This trend was also observed in the extracted kinetic parameters. As the U content increased, the saturation and maximum slope decreased, while the lag time increased. These results show that the addition of U progressively inhibits A+T assembly.

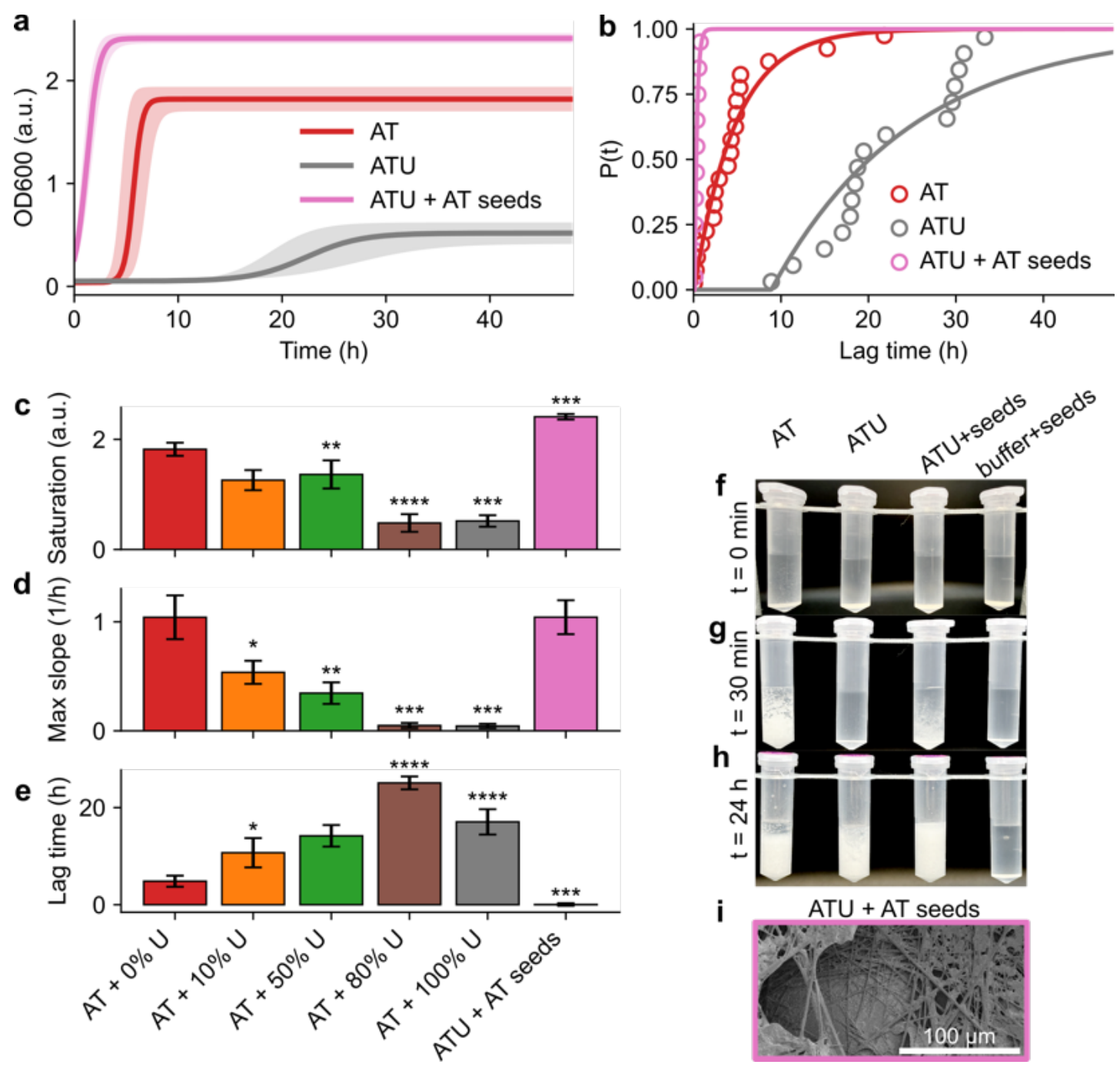

**Figure 4. Seed-assisted rescue of ternary nucleobase assembly kinetics.** (**a**) Turbidity kinetic traces and (**b**) probability of crystallization as a function of induction time for A+T, A+T+U, and A+T+U supplemented with 5% A+T seeds, measured over 48 h at 45 mM for each nucleobase component. In panel **a**, curves show averaged logistic fits of replicate measurements, with shading representing the standard error of the mean. In panel **b**, the probability function was calculated from the measured lag times of individual replicate wells. Addition of A+T seeds induces rapid assembly of the ternary mixture and shifts the crystallization probability of A+T+U to shorter induction times. (**c–e**) Fitting-derived kinetic parameters for A+T with increasing U content and for seeded A+T+U: (**c**) saturation, (**d**) maximum slope, and (**e**) lag time. Error bars represent the standard error of the mean across replicate fits. Seeding increases the saturation and maximum slope while reducing the lag time, consistent with seed-assisted rescue of inhibited A+T+U assembly. Statistical significance is indicated relative to A+T: *, $p < 0.05$; **, $p < 0.01$; ***, $p < 0.001$; ****, $p < 0.0001$. (**f–h**) Optical images of A+T, A+T+U, A+T+U with A+T seeds, and buffer with A+T seeds after (**f**) 0 min, (**g**) 30 min, and (**h**) 24 h. At 0 min, all four solutions are visually clear. After 30 min, visible assemblies are observed in A+T and seeded A+T+U, while unseeded A+T+U and buffer with A+T seeds remain mostly clear. After 24 h, unseeded A+T+U also becomes opaque. (**i**) SEM image of A+T+U assembled in the presence of A+T seeds, showing elongated crystalline fibers like those observed for the A+T binary system.

The extracted lag times from the A+T series with increasing U content, the A+U series with increasing T content, and the U+T control system were then plotted on a ternary A/T/U composition map (Figure 3b). The lag time increased toward the center of the ternary map, indicating that compositions containing substantial fractions of all three nucleobases exhibited the strongest kinetic inhibition. The shaded region marks compositions that formed distinct morphologies by SEM (Figure S3), consistent with the formation of mixed or co-assembled crystalline structures. In contrast, U+T did not form a distinct co-assembled morphology under these conditions, as shown in the Supporting Information Figure S2.

To address whether the expanded kinetic frustration observed in the ternary A/T/U mixture (following combinatorial explosion at the nucleation level) could be overcome by seeding, we tested whether preformed A+T nuclei can direct assembly in the ternary mixture. In this experiment, 5% A+T seeds were added to the A+T+U solution after the heating and cooling procedure. The seeds were taken from a pre-incubated A+T solution. Immediately after seed addition, all four solutions remained visually clear, including A+T, A+T+U, A+T+U with A+T seeds, and buffer with A+T seeds (Figure 4f). After 30 min, visible assemblies were observed in the seeded A+T+U solution, similar to the A+T solution, whereas the unseeded A+T+U solution remained mostly clear (Figure 4g). Thus, A+T seeds rapidly initiated assembly from the ternary mixture. After 24 h, the unseeded A+T+U solution also became opaque (Figure 4h), indicating that the ternary mixture was delayed rather than incapable of assembly.

The turbidity kinetics showed the same trend. Addition of A+T seeds to the A+T+U mixture led to immediate assembly (indicated by rapid turbidity development), with a kinetic trace similar to that of the A+T binary system (Figure 4a). The kinetic parameters extracted from the logistic fits showed that seeding increased both saturation and maximum slope and significantly reduced the lag time (Figure 4c–e). This corresponds to affecting both the nucleation and the growth rate of the crystals. To further quantify the nucleation effect, the extracted lag-time distributions were used to determine the apparent nucleation rate using classical nucleation theory (CNT).[39] Lag-time distributions were converted to cumulative crystallization probability functions and fitted to an exponential onset model to extract the nucleation rate, $J$, using the following function:

$$P(t) = 1 - e^{-JV(t_{lag} - t_g)}$$

Where $J$ is the nucleation rate, $V$ is the sample's volume, $t_{lag}$ is the extracted lag time, and $t_g$ is the time between the formation of critical nuclei and their growth to measurable turbidity. This analysis, presented in Figure 4b, showed that the ternary A+T+U mixture had a lower $J$ than the binary A+T system, 115±16 compared with 397± 28 $[m^{-3}s^{-1}]$, whereas addition of A+T seeds increased J to 6710± 590 $[m^{-3}s^{-1}]$. The full fitting parameters and fits for different concentrations of U are presented in Figure S8 and Table S4. SEM imaging of the seeded A+T+U sample showed elongated crystalline fibers, similar to those observed for A+T, rather than the largely disordered aggregates observed for unseeded A+T+U after 48 h (Figure 4h).

To directly determine the chemical composition of the crystalline fraction, crystals were collected by filtration, washed, redissolved, and analyzed by HPLC at 254 nm (Figure 5). In contrast to the turbidity measurements, which probe the full suspension, this analysis selectively examined the solid material formed in each sample. Crystals from the binary A+T system contained 66 ± 2% A, 34 ± 2% T, and 0.45 ± 0.15% U in molar fraction. Crystals from the ternary A+T+U system contained 48 ± 2% A, 33 ± 2% T, and 19 ± 1% U. In the A+T+U sample seeded with A+T crystals, the crystalline fraction shifted toward an A/T-rich composition, with 66 ± 11% A, 32 ± 11% T, and 2.1 ± 0.9% U. Quantification was performed using standard solutions, as detailed in Figure S13 and Table S6.

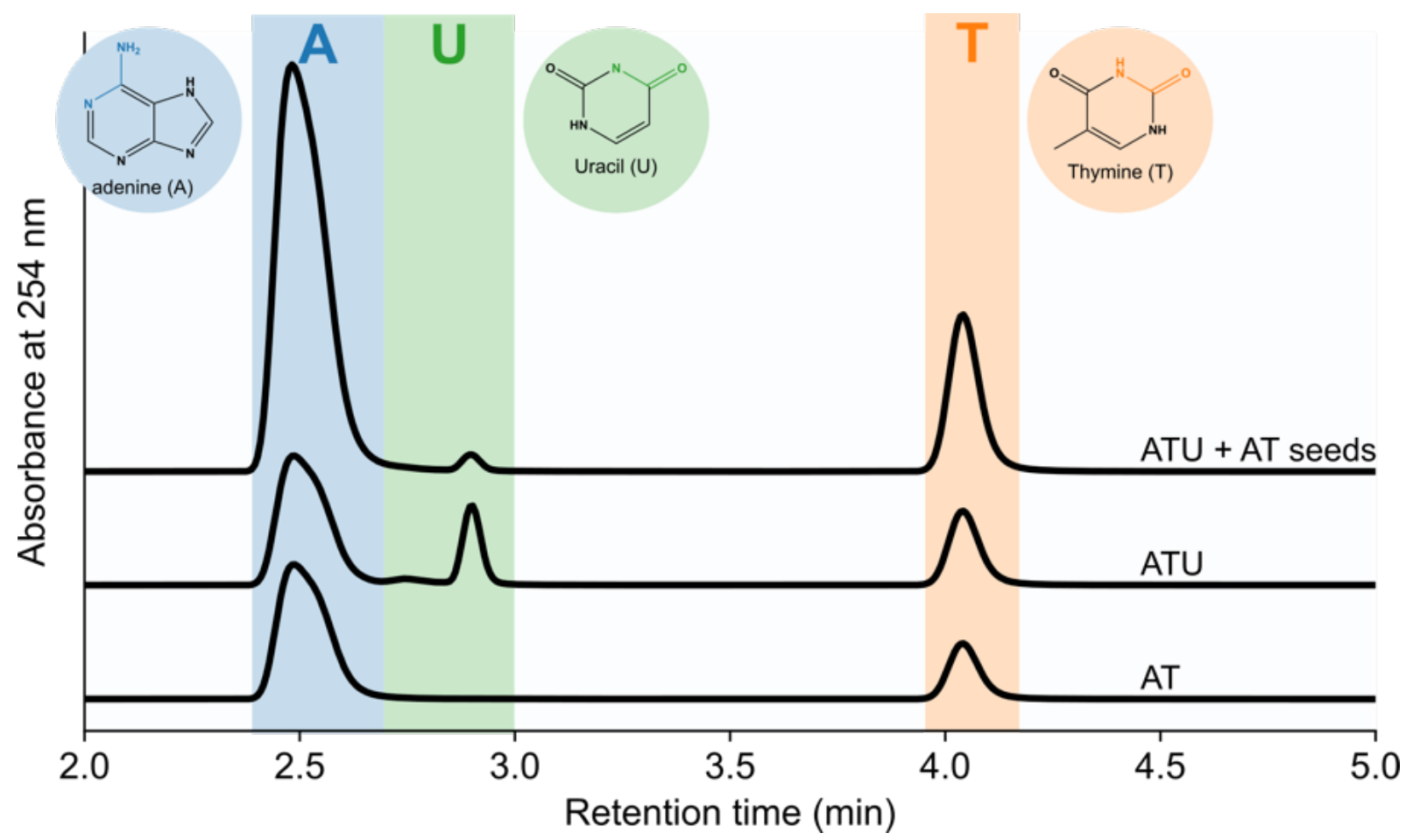

**Figure 5. HPLC analysis of redissolved crystals.** HPLC chromatograms at 254 nm of collected and redissolved crystals from A+T, A+T+U, and A+T+U seeded with A+T crystals, shown from bottom to top. A and T are detected in the binary and seeded systems, while A, T, and U are detected in the ternary system. Retention times of A, U, and T are marked in blue, green, and orange, respectively.

## Discussion

Here, we show that ternary nucleobase crystallization is kinetically frustrated by competition among closely related hydrogen-bonding motifs, and that crystalline seeds restore rapid growth by selectively channeling assembly into a defined ordered pathway. Compared with the A+T binary system, the A+T+U ternary mixture showed a pronounced delay in assembly, with longer lag times and a lower apparent nucleation rate. This kinetic signature indicates that the addition of U primarily affects the early stages of crystallization rather than simply changing the final solid-state morphology. Because U is a canonical nucleobase with hydrogen-bonding functionality closely related to T, and nucleobases are known to support multiple canonical and noncanonical recognition modes,[21] U can plausibly introduce additional local recognition motifs and competing early assembly states. We therefore interpret the delayed crystallization of A+T+U as competition-induced kinetic frustration: the ternary mixture is not incapable of ordered assembly, but its access to the A+T-like crystalline pathway is delayed by competition among multiple possible nucleation trajectories.

This kinetic frustration could be overcome by introducing preformed A+T seeds. A small amount of A+T seeds sharply reduced the lag time and increased the apparent nucleation rate of A+T+U, consistent with the established role of seeding in bypassing nucleation barriers and directing crystallization or self-assembly outcomes.[16–18] Importantly, seeding did not merely accelerate the appearance of any solid phase. Product analysis by HPLC showed that the seeded ternary mixture was biased toward the A+T crystalline product, with reduced U incorporation relative to the unseeded A+T+U assembly. Thus, the A+T seed acts as a selector of molecular composition as well as assembly pathway: it provides a preorganized crystalline interface that favors growth through A+T interactions in a solution that also contains a competing A-recognition partner.

These results place the A/T/U system as a simple model for pathway selection in chemically competitive supramolecular assembly within, but distinct from, established examples of additive-controlled and competitive crystallization. Structurally related additives are known to inhibit or reshape crystallization through specific intermolecular interactions, including modification of crystal morphology and growth kinetics.[6,8,10,11] Competitive cocrystallization can also select among related molecular components when different possible cocrystals form with different kinetics.[15] The A/T/U system is a complementary and chemically simplified model case in which three canonical nucleobases compete through closely related hydrogen-bonding recognition motifs in water, and seeding biases the molecular subset incorporated into the crystalline product. This provides a minimal model for seeded compositional selection, in which a crystalline interface selects not only when crystallization occurs, but also which molecular partner is amplified from a competitive supramolecular mixture.

Combinatorial compression reflects the selective narrowing of a large chemical possibility space into a smaller number of persistent products or configurations. In prior work, chemical selection has been observed across multiple scales and through various mechanisms. Wet-dry cycling uses kinetic control and energy dissipation to select for complex oligomers.[34,40] Mineral evolution selects for locally stable arrangements of chemical elements.[41] These phenomena may be explained in part by broader physical frameworks for selection in driven systems, such as Bejan's theory of Constructal Law[42–44] or England's theory.[45] Consistent with this view, our recent work has demonstrated that in wet-dry cycling of complex prebiotic mixtures, small kinetic advantages can strongly select a narrow set of condensation products, producing combinatorial compression rather than combinatorial explosion.[34,35] Moreover, competition among self-replicating molecules can produce temporal compression, meaning that a small set of replicators dominates early and curtails chemical diversification over time, thereby limiting the system's tendency to spread into many products.[46] This view is consistent with recent proposals that evolving systems, including nonbiological systems, emerge when many possible configurations are generated but only a subset is preferentially preserved because it performs a function such as persistence, propagation, or novelty generation.[47] The present work extends this idea from covalent product distributions to noncovalent assembly landscapes. In the A/T/U system, the molecular components are not transformed into new covalent products. Instead, compression occurs through differential access to and amplification of a limited crystalline state from a larger set of possible supramolecular arrangements.

A central challenge in prebiotic chemistry is therefore not only how biologically relevant molecules were formed, but how specific molecular subsets could be selected and preserved from chemically complex mixtures. Extraterrestrial delivery and prebiotic synthesis can generate diverse mixtures of canonical and noncanonical nucleobases, nucleosides, and related heterocycles.[48,49] In such mixtures, crystallization can act as a physical selection mechanism by enriching a subset of molecules from a broader chemical pool, as shown for prebiotic intermediates such as ribose aminooxazoline and for nucleotide precursors under evaporative conditions.[28,50] Our results extend this idea to competitive nucleobase self-assembly: in the ternary A/T/U system, mixture complexity suppresses the formation of the ordered A+T crystalline phase relative to the binary system, demonstrating that additional components can frustrate assembly. However, once A+T seeds are introduced, the ternary mixture is redirected toward the faster-forming crystalline state. Thus, a selected crystalline phase can act as both a reservoir of selected molecules and a selector of molecular composition for future assembly, while molecules that remain in solution may be diluted, washed away, degraded, or recycled. Thus, seeded crystallization provides a simple physical route by which chemically complex environments could become compositionally organized before the emergence of biological replication.

## Conclusion

This work establishes seeded crystallization as a mechanism for compositional pathway selection in a complex nucleobase mixture. The A+T+U ternary system is not incapable of ordered assembly but is kinetically frustrated by competition among related hydrogen-bonding recognition motifs, leading to delayed and less efficient crystallization. Preformed A+T seeds overcome this frustration by restoring rapid growth and biasing the solid product toward the A+T crystalline pathway. We describe this process as kinetic compression: a pre-existing ordered phase reduces the accessible supramolecular assembly landscape by amplifying one selected molecular trajectory from a multicomponent solution. More broadly, these results suggest that crystallization could provide a simple physical route for selecting and preserving molecular subsets from chemically complex environments.

## Materials

Adenine, thymine, and uracil were purchased from Aaron Chemicals (adenine, AR00354N; thymine, AR003VI3; uracil, AR0036HB) and used as received. A 0.2 M phosphate buffer was prepared analytically with sodium dihydrogen phosphate monohydrate in triple-distilled water using a Milli-Q SQ 2 Series benchtop water purification system. The buffer was adjusted to pH 7.0 using 1 M NaOH and stored at 4 °C until use. Unless otherwise stated, all solutions were prepared using the phosphate buffer.

## Crystallization by supersaturation

Crystallization experiments were performed under supersaturation conditions. The required amount of each nucleobase was weighed into a Falcon tube, followed by the addition of buffer to the desired final concentration. For standard experiments, solutions were prepared at a final concentration of 45 mM for each component in a total volume of 10 mL. The solutions were heated at 95 °C for at least 3 h with intermittent mixing via vortex until complete dissolution. After dissolution, the solutions were allowed to cool for several minutes until they could be safely pipetted.

For kinetic measurements, the warm solutions were transferred directly into the plate reader plate as described below. For crystallization samples intended for microscopy, HPLC analysis, FTIR, and EDX (see Figures S11- 12), the solutions were left in Falcon tubes to cool and crystallize under ambient conditions. Turbidity and visible crystal formation were typically observed within approximately 24 h, and samples were generally incubated for 48 h before further analysis.

**Plate reader kinetic measurements**

Crystallization kinetics were monitored by optical density measurements at 600 nm using a SYNERGY H1 microplate reader (Biotek). Different wavelengths were tried and shown not to affect the measurement precision (Figure S7). For each experimental condition, at least 8 replicate wells from the same solution were prepared, with a final volume of 150 µL per well. A flat-bottom 96-well plate was sealed using MicroAMP Optical Adhesive Film by applied biosystems; thermofisher. Turbidity was measured every 10 min for 48 h at 30 °C.

Before the first measurement, the plate was mixed by double-orbital shaking at 425 rpm for 10 s. Before each subsequent measurement, the plate was gently mixed at 180 rpm for 5 s. Each well was treated as an independent replicate.

For the seeding experiments, a pre-prepared solution of AT crystals was used for seeding. 7.5 µL of the crystalline solution was added to the 150 µL well of the supersaturated solution, corresponding to a 5% molar ratio addition. The seeded solution was then mixed and measured using the same protocol.

**Kinetic analysis**

The resulting kinetic traces were fitted using a logistic function:

$$y = b + \frac{A}{1 + e^{-k(t-t_0)}}$$

The following kinetic parameters were extracted for each replicate well: saturation ($S = b + A$); max slope ($s_{max} = \frac{Ak}{4}$); and lag time ($t_{lag} = t_0 - \frac{2}{k}$). For each condition, individual replicate traces were fitted separately. The extracted fitting parameters were then averaged across replicates, and the uncertainty is reported as the standard error of the mean. For visualization of the kinetic traces, the mean fitted curve is shown with the standard error of the mean represented as the shaded region.

Statistical significance was determined using a two-tailed Student's t-test assuming unequal variance. Differences were considered significant at $p < 0.05$. Full pairwise comparisons and corresponding p-values are provided in the Supporting Information.

The extracted lag time distributions were used to extract the apparent nucleation rate using classical nucleation theory (CNT)[39]. The extracted lag times were sorted in ascending order and converted to a cumulative crystallization probability using $P_i = (i - 0.5)/N$, where $i$ is the rank of each lag time and ($N$) is the total number of replicate wells. The resulting cumulative probability distribution was fitted to an exponential Poisson onset model: $P(t) = 1 - e^{-JV(t_{lag} - t_g)}$. Where $J$ is the nucleation rate, $V$ is the sample's volume, $t_{lag}$ is the extracted lag time, and $t_g$ is the growth time to measurable turbidity. The fit returned $JV$ and $t_g$; $J$ was then calculated by dividing $JV$ by the corresponding well volume. The uncertainty in $J$ was calculated by error propagation from the fitted uncertainty in $JV$. Because turbidity measurements report the time required for assemblies to reach an optical detection threshold rather than the appearance of molecular nuclei directly, and because the experiments were performed in small 96-well-plate volumes, the extracted $J$ values should be interpreted as comparative apparent nucleation/onset rates under these measurement conditions. The analysis and figure preparation were done in Python using Spyder6.

**Scanning electron microscopy**

Scanning electron microscopy samples were prepared on silicon wafers. The wafers were cleaned sequentially with acetone and ethanol and then dried using pressurized air. A small droplet of each crystallization sample was deposited onto the silicon wafer and allowed to dry under ambient conditions.

To reduce charging during imaging, the dried samples were coated with iridium using Quorum Q150VS sputterer. SEM imaging was performed using an Extra-High-Resolution Scanning Electron Microscope, Magellan 400L (Thermo Fisher Scientific). Images were acquired using an accelerating voltage of 2 kV and a beam current of 6.3 pA.

**HPLC**

Crystalline products were prepared in triple-distilled water produced using a Milli-Q SQ 2 Series benchtop water purification system. The resulting crystals were collected and washed by Büchner filtration and redissolved in triple-distilled water to a final concentration of 0.2 mg $mL^{-1}$ before analysis. The sample was injected at 0.2 and 0.1 mg/mL. Quantification was performed using 0.2mg analytical-grade standards prepared under the same solvent conditions. The standards were diluted to various concentrations (0.01-0.2 mg/mL) and measured in the HPLC to create a calibration curve (See Figure S13)

HPLC analysis was carried out on an Agilent Infinity II 1260 system equipped with a PDA detector. Separation was performed using an Agilent InfinityLab Poroshell EC-C18 column equipped with a C18 guard column, 4 × 2 mm. The column temperature was maintained at 30 °C. Samples were analyzed using an isocratic method with a mobile phase composed of 95% solvent A and 5% solvent B, where solvent A was water containing 0.1% formic acid and solvent B was acetonitrile. The run time was 5 min, the injection volume was 5 µL, and detection was performed at 254 nm.

## Funding

This research was funded by the European Union (ERC, Sweet_Evo, 101163270). Views and opinions expressed are, however, those of the authors only and do not necessarily reflect those of the European Union or the European Research Council. Neither the European Union nor the granting authority can be held responsible for them.

# Supporting Information

# Combinatorial Compression in Nucleobase Self-Assembly Governed by Seeding Kinetics

Yael Kapon[a], Rotem Edri[a], and Moran Frenkel-Pinter[a,b,c*]

[a] Institute of Chemistry, The Hebrew University of Jerusalem, Jerusalem, Israel

[b] The Center for Nanoscience and Nanotechnology, The Hebrew University of Jerusalem, Jerusalem, Israel 9190401

[c] Casali Center of Applied Chemistry, The Hebrew University of Jerusalem, Jerusalem 9190401, Israel

*Corresponding author:

Dr. Moran Frenkel-Pinter
The Center for Nanoscience and Nanotechnology
Institute of Chemistry
The Hebrew University of Jerusalem
Edmond J. Safra Campus
Jerusalem 9190401, Israel
Ph: (+972)-2-6584171
moran.fp@mail.huji.ac.il

## Table of Contents

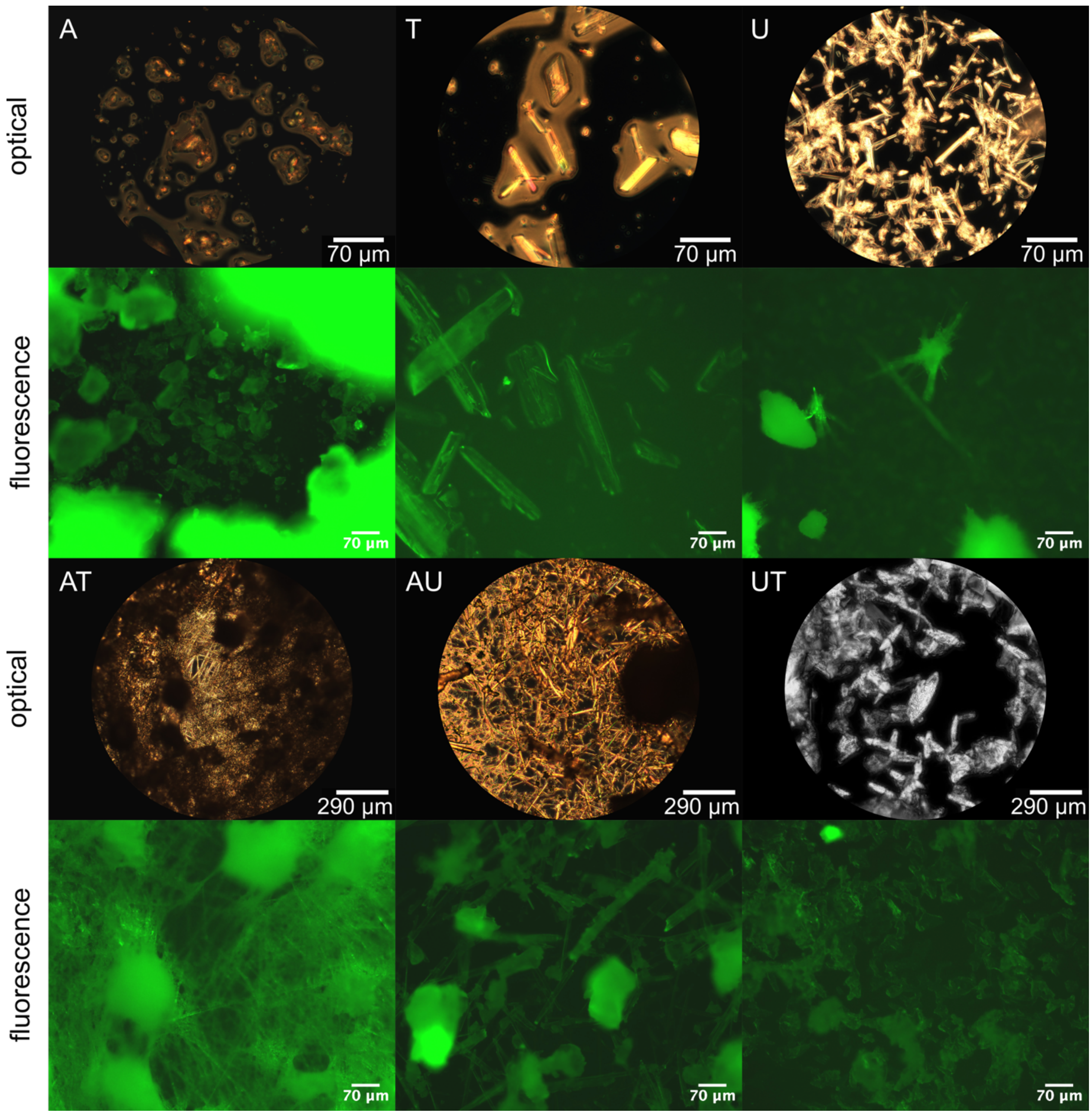

**Figure S1. Optical microscopy images of the nucleobase structures.** Cross-polarized optical (top) and fluorescence (bottom) microscopy images of single component and binary nucleobase systems. All systems formed defined crystal morphologies and exhibited autofluorescence in the green channel.

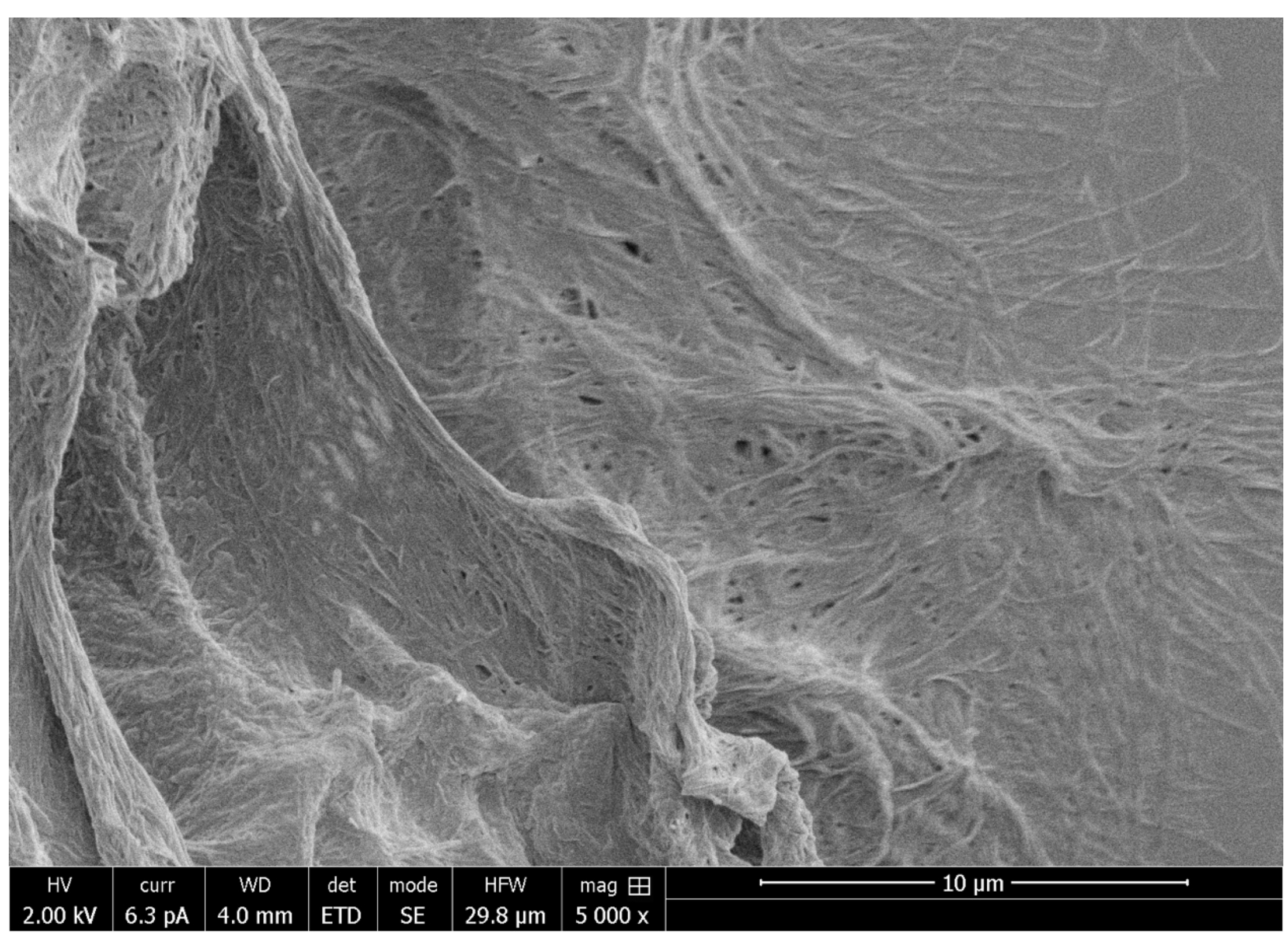

**Figure S2: High-resolution SEM of the A+T+U system showing elongated fibers**

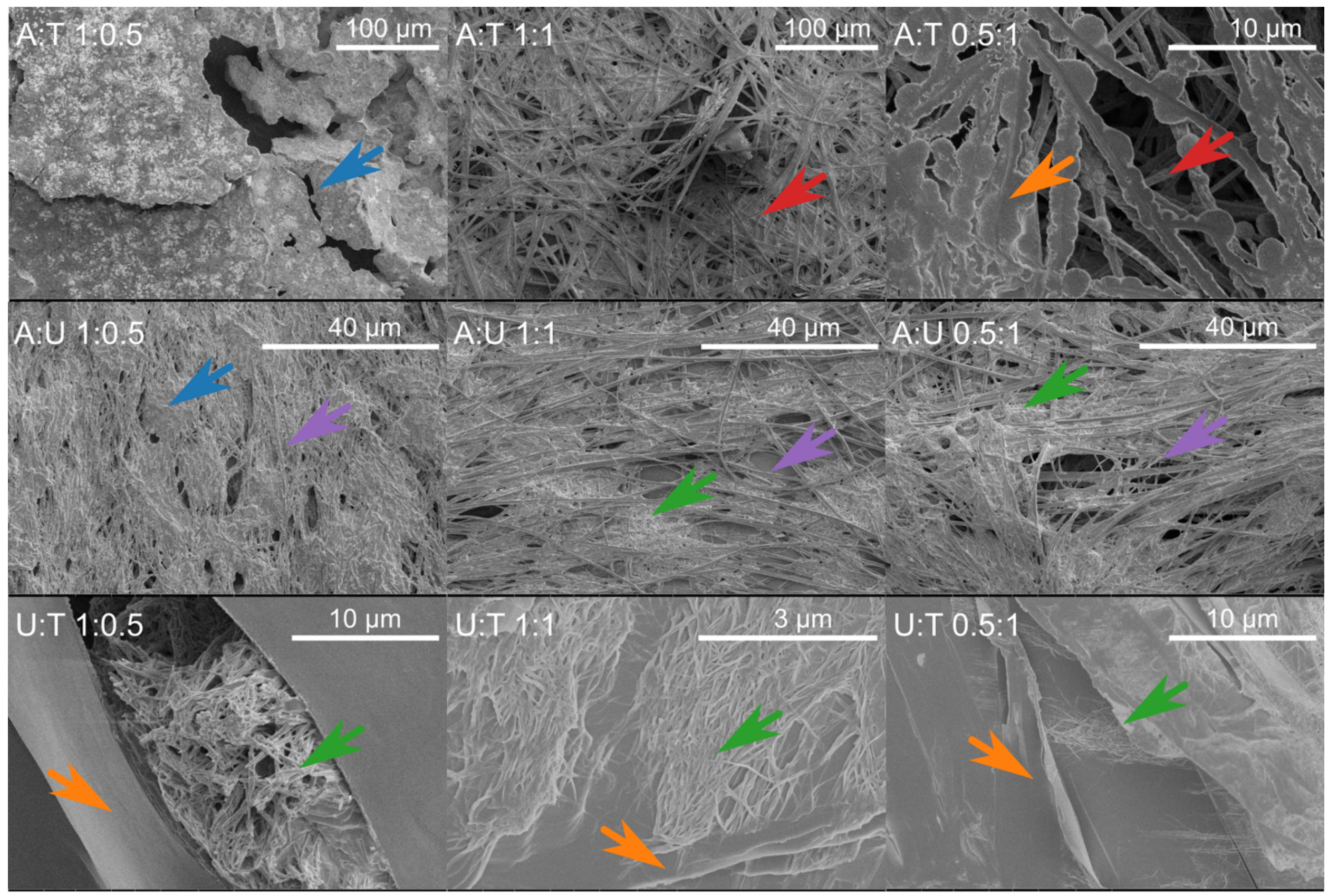

**Figure S3. Morphological comparison of binary nucleobase assemblies at different molar ratios.** SEM images of samples with different ratios of nucleobases showing either single-component morphology (A-blue; T-orange; U-green) or binary co-assembly morphology (A+T-red; A+U-purple). U+T did not show any distinct binary co-assembly morphology.

**Kinetic analysis workflow and fitting parameters**

Kinetic traces were measured using turbidity +600 nm in a plate reader, where each well was treated as an independent replicate. Figure S3 shows all raw kinetic traces measured for A+T(45 mM each component) and A+T+U (45 mM each component), plotted in red and gray, respectively. These traces show the well-to-well variability expected for crystallization processes, where nucleation is stochastic, and each replicate may initiate growth at a slightly different time.

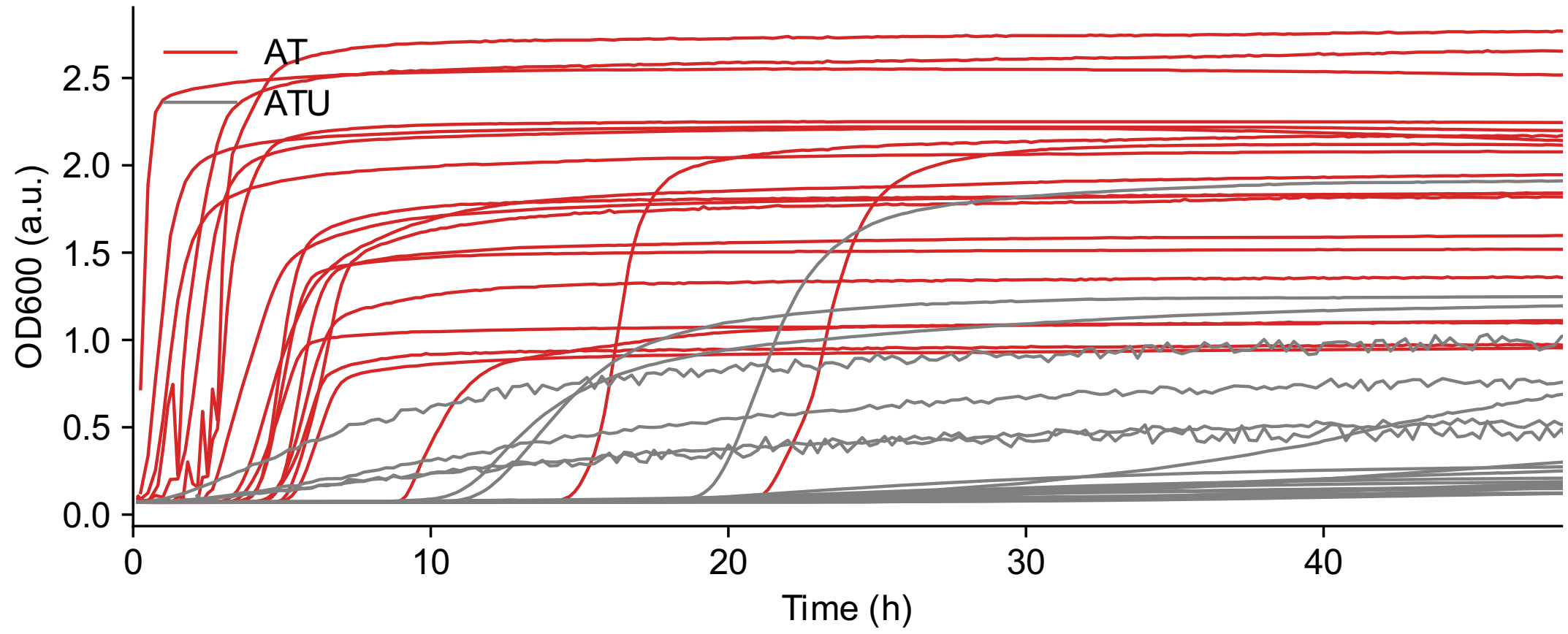

**Figure S4.** Raw kinetic traces for individual replicate wells of A+T and A+T+U, each prepared at 45 mM per nucleobase component, are shown in red and gray, respectively.

The average kinetic traces are shown in Figure S4, with the shaded region representing the standard error of the mean. The averaged traces show step-like features. These features arise from averaging multiple sigmoidal traces with different lag times and growth rates, rather than from discrete steps in a single crystallization event, reflecting the statistical nature of nucleation and growth across replicate wells.

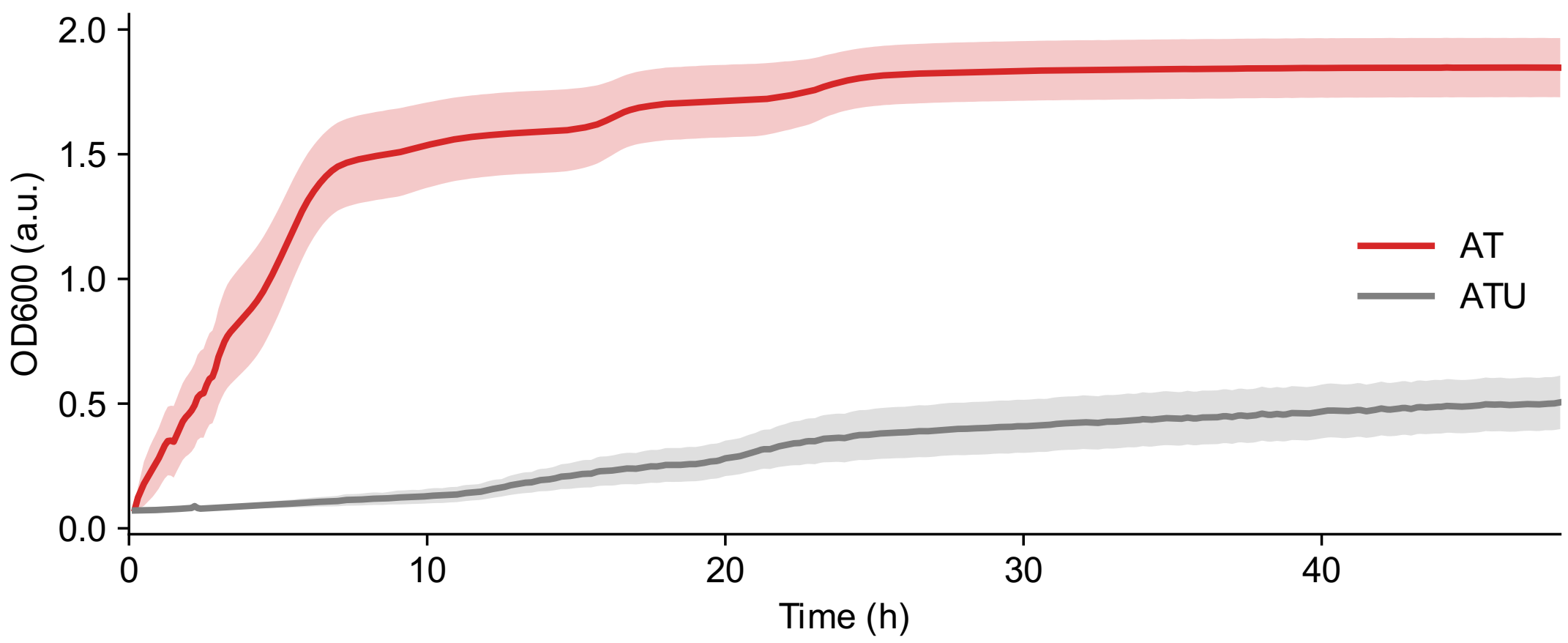

**Figure S5.** Average kinetic traces for A+T and ATU, each prepared at 45 mM per nucleobase component, with the standard error of the mean shown as shaded regions.

To extract comparable kinetic parameters while avoiding artifacts introduced by direct averaging, each individual replicate trace was fitted separately using a logistic function. The fitted parameters were then averaged across replicates. The logistic fit provides three main descriptors of the crystallization kinetics: lag time, maximum slope, and saturation OD. Representative fits for A+T and A+U are shown in Figure S5, top and bottom, respectively.

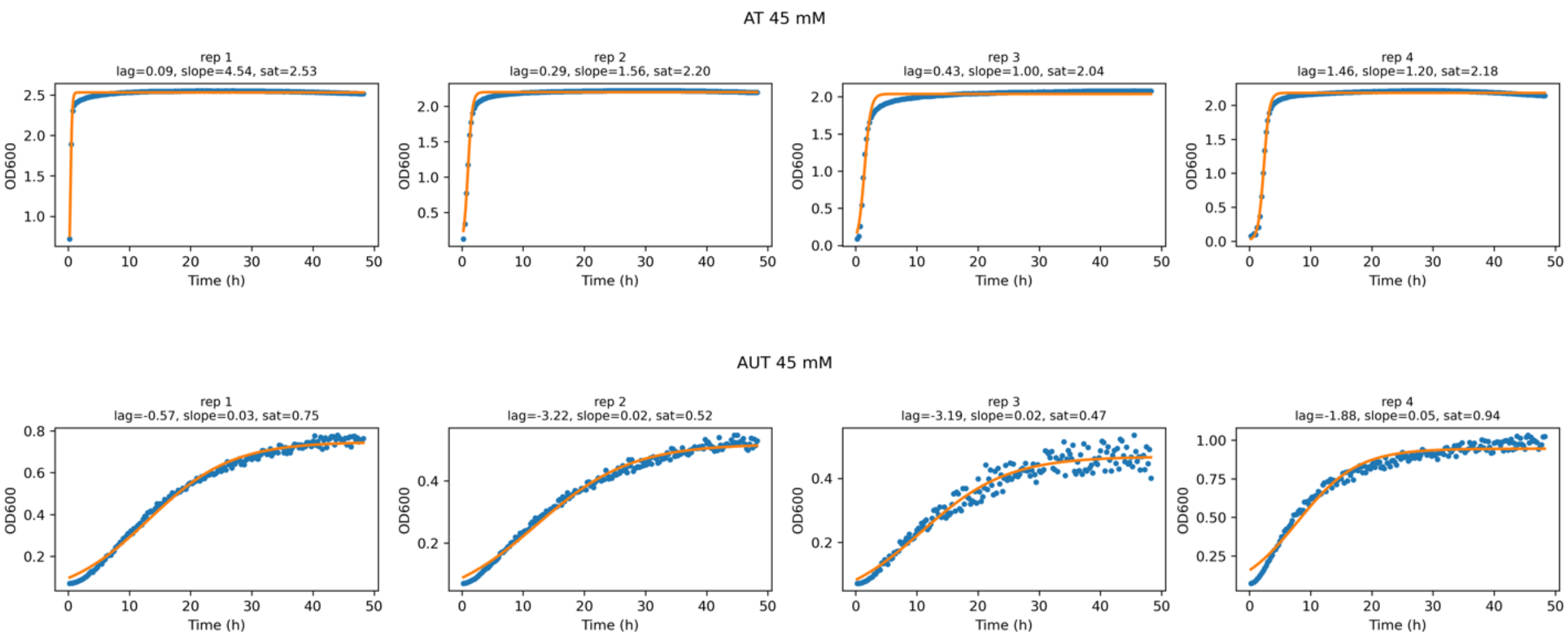

**Figure S6.** Examples of logistic fitting to individual kinetic traces for A+T (top) and A+U (bottom) crystallization.

For visualization in Figures 2, 3, and 4, an averaged sigmoid was reconstructed using the mean fitted parameters for each condition. The uncertainty region was estimated from the standard error of the mean (SEM) of the fitted parameters. The upper boundary was generated using lag time − SEM, saturation + SEM, and maximum slope + SEM, while the lower boundary was generated using lag time + SEM, saturation − SEM, and maximum slope − SEM. This approach provides a visual estimate of the uncertainty in the fitted kinetic behavior while preserving the sigmoidal shape of the crystallization process.

**Statistical analysis of fitted kinetic parameters**

Pairwise statistical comparisons between fitted kinetic parameters were performed for the A+T + %U composition series, including the seeded ATU sample. For each condition, only successful sigmoid fits were included. Statistical significance was determined using a two-tailed Student's t-test assuming unequal variance, equivalent to Welch's t-test. The resulting pairwise p-values are reported below for lag time, maximum slope, and saturation. Values reported as <0.0001 correspond to values that were rounded to 0.0000 in the analysis output.

The following color map corresponds to the different p-values:

| p-value | $p \geq 0.05$ | $p < 0.05$ | $p < 0.01$ | $p < 0.001$ | $p < 0.0001$ |
|---|---|---|---|---|---|
| significance | NS | * | ** | *** | **** |

Table S1. Pairwise p-values for fitted lag time.

| **Condition** | **A+T** | **A+T + 10% U** | **A+T + 50% U** | **A+T + 80% U** | **A+T+U** | **A+T+U + A+T seeds** |
|---|---|---|---|---|---|---|
| **A+T** | — | 0.1222 | 0.0050 | <0.0001 | 0.0003 | 0.0007 |
| **A+T + 10% U** | 0.1222 | — | 0.4004 | 0.0024 | 0.1479 | 0.0130 |
| **A+T + 50% U** | 0.0050 | 0.4004 | — | 0.0021 | 0.4300 | 0.0005 |
| **A+T + 80% U** | <0.0001 | 0.0024 | 0.0021 | — | 0.0137 | <0.0001 |
| **A+T+U** | 0.0003 | 0.1479 | 0.4300 | 0.0137 | — | <0.0001 |
| **A+T+U + A+T seeds** | 0.0007 | 0.0130 | 0.0005 | <0.0001 | <0.0001 | — |

Table S2. Pairwise p-values for fitted maximum slope.

| **Condition** | **A+T** | **A+T + 10% U** | **A+T + 50% U** | **A+T + 80% U** | **A+T+U** | **A+T+U + A+T seeds** |
|---|---|---|---|---|---|---|
| **A+T** | — | 0.0412 | 0.0059 | 0.0001 | 0.0001 | 0.9976 |
| **A+T + 10% U** | 0.0412 | — | 0.2378 | 0.0032 | 0.0031 | 0.0202 |
| **A+T + 50% U** | 0.0059 | 0.2378 | — | 0.0264 | 0.0245 | 0.0022 |
| **A+T + 80% U** | 0.0001 | 0.0032 | 0.0264 | — | 0.8855 | 0.0001 |
| **A+T+U** | 0.0001 | 0.0031 | 0.0245 | 0.8855 | — | 0.0001 |
| **A+T+U + A+T seeds** | 0.9976 | 0.0202 | 0.0022 | 0.0001 | 0.0001 | — |

Table S3. Pairwise p-values for fitted saturation.

| **Condition** | **A+T** | **A+T + 10% U** | **A+T + 50% U** | **A+T + 80% U** | **A+T+U** | **A+T+U + A+T seeds** |
|---|---|---|---|---|---|---|
| **A+T** | — | 0.0301 | 0.1573 | <0.0001 | <0.0001 | 0.0002 |
| **A+T + 10% U** | 0.0301 | — | 0.7623 | 0.0097 | 0.0063 | 0.0005 |
| **A+T + 50% U** | 0.1573 | 0.7623 | — | 0.0183 | 0.0176 | 0.0061 |
| **A+T + 80% U** | <0.0001 | 0.0097 | 0.0183 | — | 0.8591 | <0.0001 |
| **A+T+U** | <0.0001 | 0.0063 | 0.0176 | 0.8591 | — | <0.0001 |
| **A+T+U + A+T seeds** | 0.0002 | 0.0005 | 0.0061 | <0.0001 | <0.0001 | — |

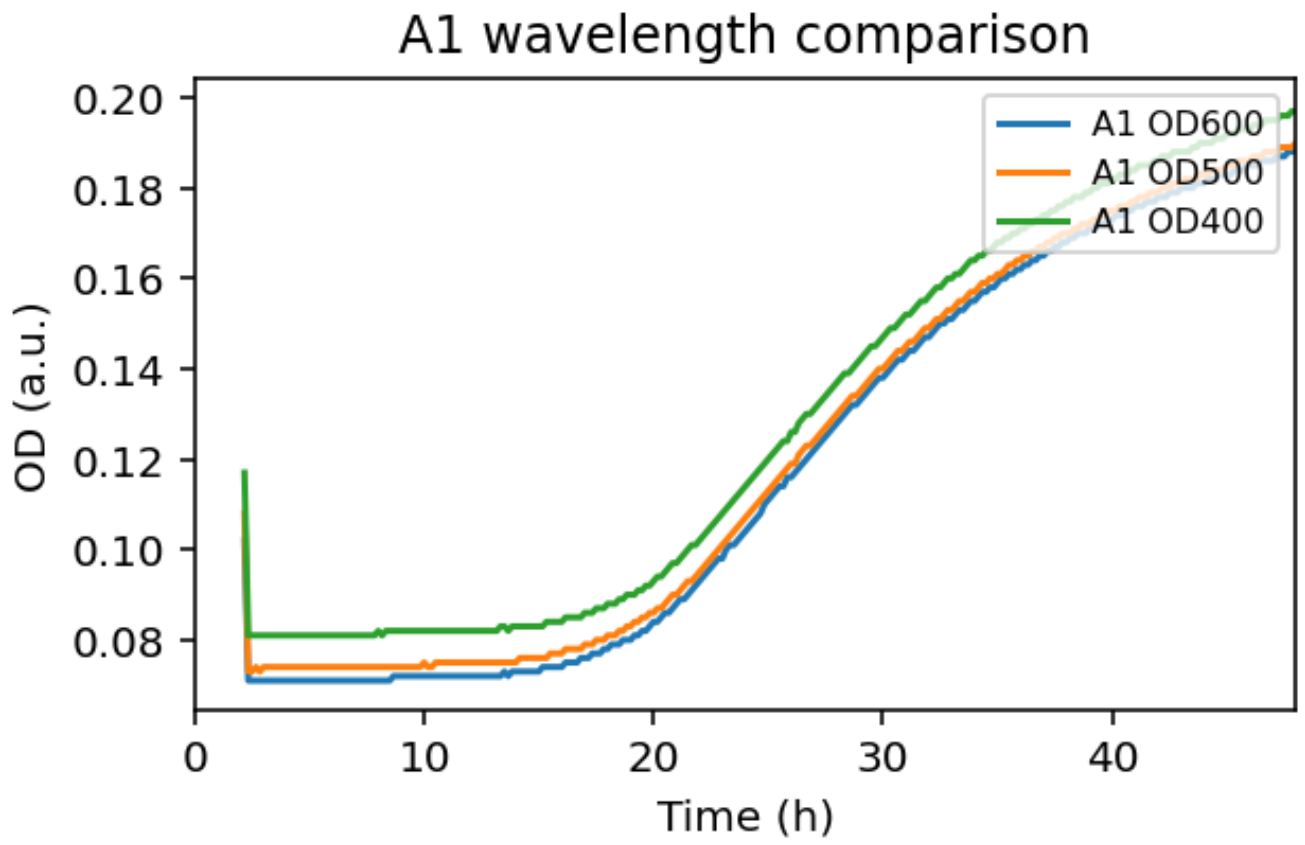

**Figure S7. Comparison of turbidity kinetics measured at different wavelengths.** Representative OD traces collected at 400, 500, and 600 nm for selected nucleobase assembly conditions show similar behavior and signal intensity. 600 nm was chosen for simplicity.

**Apparent nucleation-rate analysis from lag-time distributions**

Apparent onset rates were estimated from the cumulative probability distributions of fitted lag times. For each condition, successful sigmoid-derived lag times were sorted and converted to cumulative probabilities using $P_i = \frac{i-0.5}{N}$, where $i$ is the rank and $N$ is the number of successful fits. The probability distributions were fitted to $\mathrm{P(t)} = 1 - \frac{1}{\mathrm{e}^{-\mathrm{JV(t-t_g)}}}$, where JV is the apparent onset rate per well and $t_g$ is the growth-to-detection delay. The apparent volumetric onset rate was calculated as $J_{app} = \frac{JV}{V*3600}$, where V is the sample volume in $m^3$.

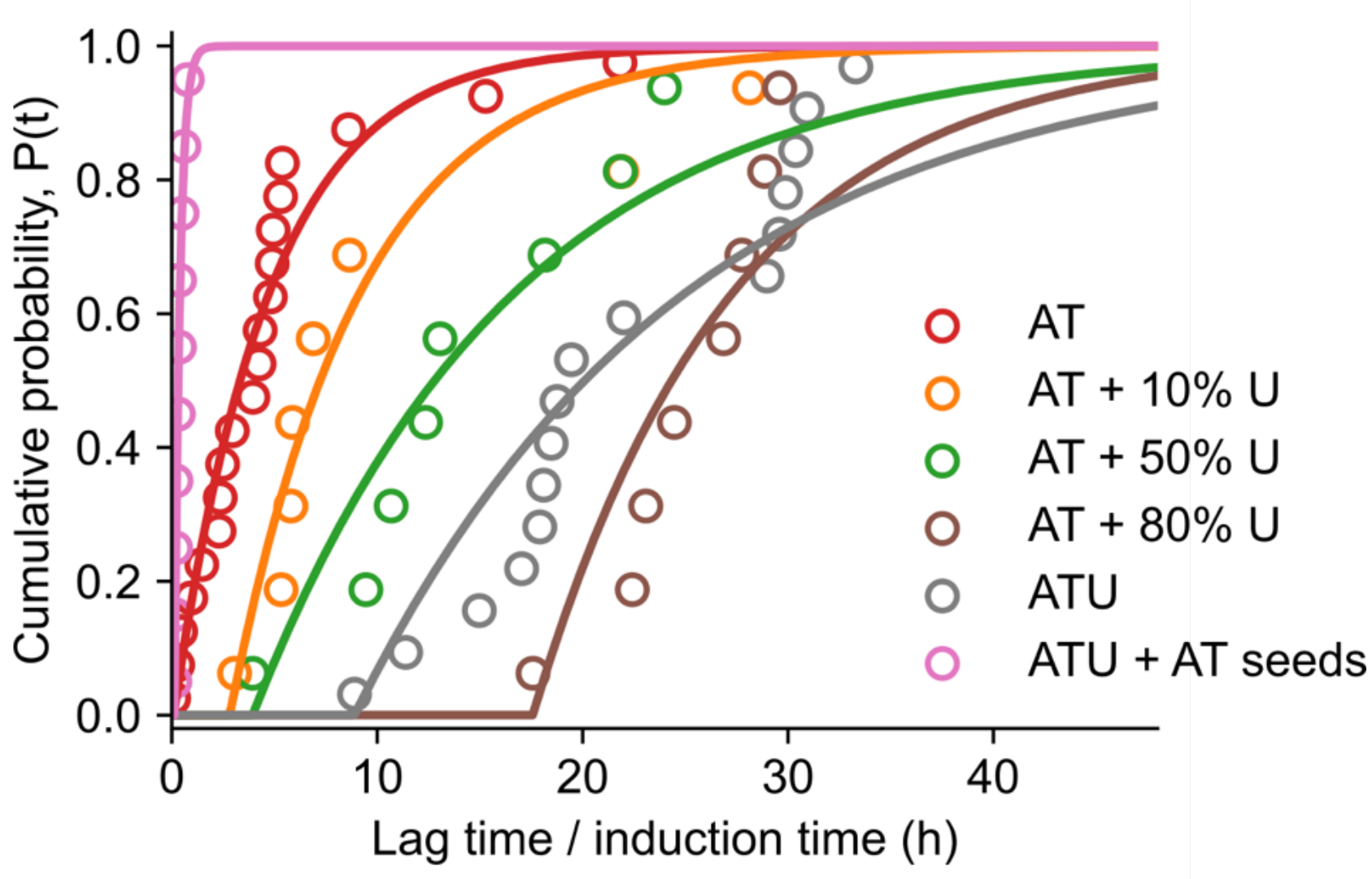

**Figure S8. Probability of crystallization as a function of induction time** for the A+T system with increasing U content measured over 48 h (45 mM A and 45 mM T, with U added at the indicated concentrations). The probability function was calculated from the measured lag times of individual replicate wells. Addition of U slows down the assembly, whereas A+T seeds induce rapid assembly of the ternary mixture and shift the crystallization probability of A+T+U to shorter induction times.

**Table S4.** Apparent onset-rate parameters extracted from lag-time probability distributions for the A+T + %U composition series.

| **Condition** | **N** | **Volume (μL)** | **JV ($h^{-1}$)** | **tg (h)** | **Japp ($m^{-3}$ $s^{-1}$)** |
|---|---|---|---|---|---|
| A+T | 20 | 150 | 0.215 ± 0.015 | 0.085 ± 0.163 | 397 ± 28 |
| A+T + 10% U | 8 | 150 | 0.158 ± 0.036 | 2.816 ± 0.652 | 292 ± 67 |
| A+T + 50% U | 8 | 150 | 0.078 ± 0.014 | 3.928 ± 1.243 | 145 ± 25 |
| A+T + 80% U | 8 | 150 | 0.102 ± 0.024 | 17.565 ± 1.384 | 190 ± 45 |
| A+T+U | 16 | 150 | 0.062 ± 0.009 | 8.885 ± 1.299 | 115 ± 16 |
| A+T+U + A+T seeds | 10 | 157.5 | 3.805 ± 0.335 | 0.127 ± 0.011 | 6711 ± 590 |

**A+U + %T system**

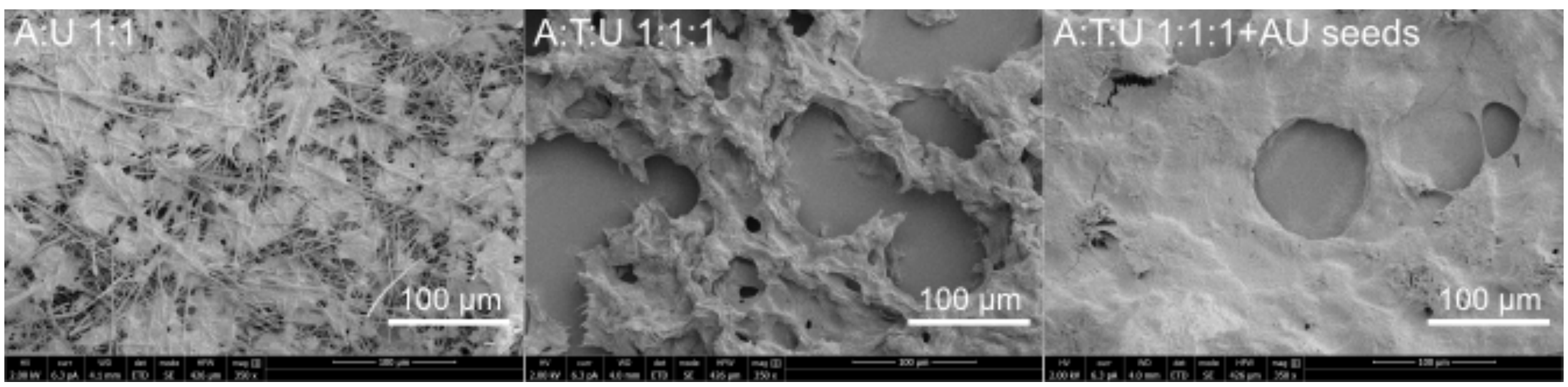

**Figure S9.** SEM images of A+U, A+T+U, and A+T+U seeded with 5% A+U seeds.

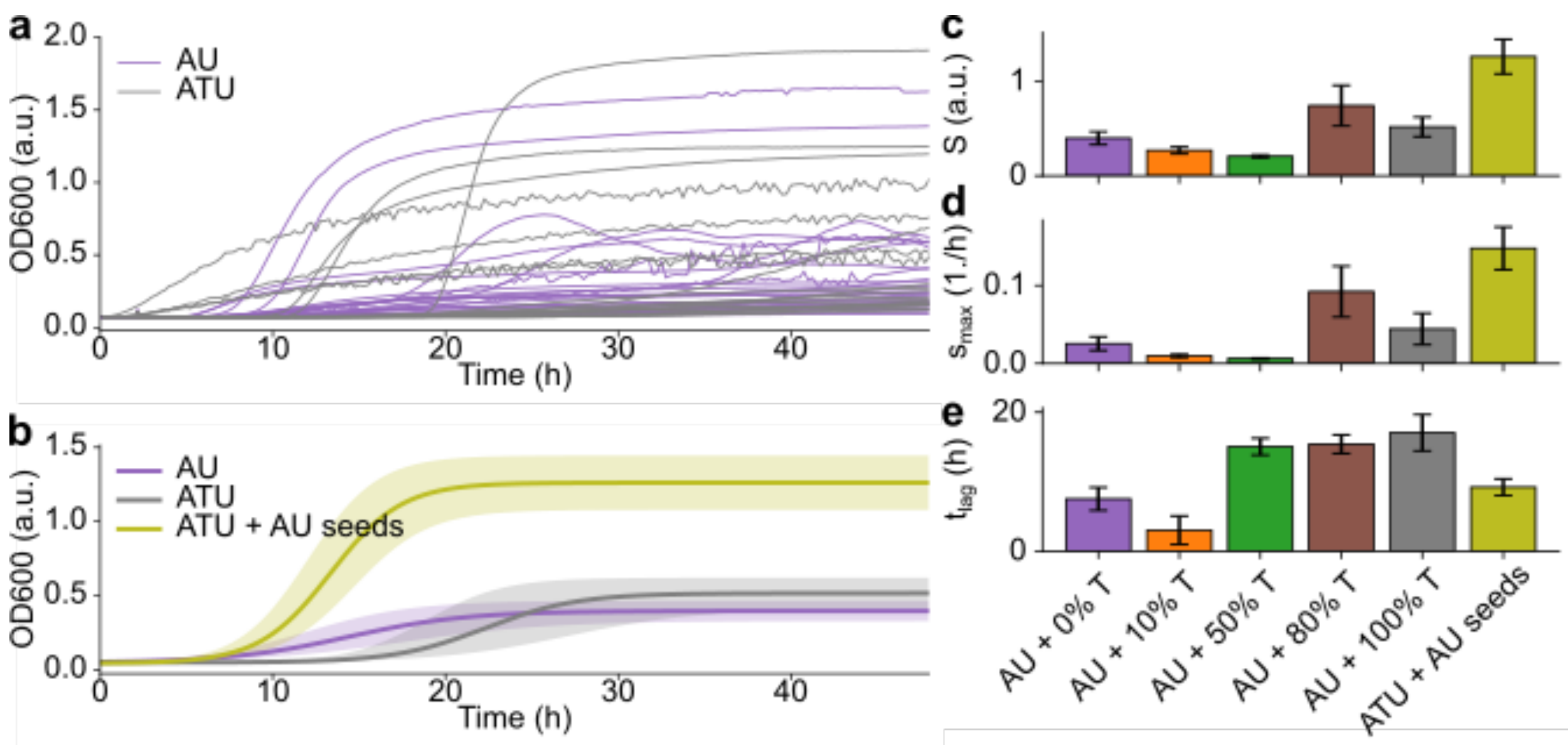

**Figure S10. Turbidity measurements of A+U, A+T+U, and A+T+U seeded with 5% A+U seeds. (a)** Raw OD600 kinetic traces and **(b)** sigmoid fits of A+U, A+T+U, and A+T+U supplemented with 5% A+U seeds, measured over 48 h at 45 mM per nucleobase component. Curves in **b** show the averaged sigmoid fits of replicate measurements, with shading representing the standard error of the mean. The addition of A+U seeds induces rapid assembly of the ternary mixture. **(c–e)** Fitting-derived kinetic parameters for A+U with increasing T content and for seeded A+T+U: **(c)** saturation, **(d)** maximum slope, and **(e)** lag time. Error bars represent the standard error of the mean across replicate fits. Seeding increases the saturation and maximum slope while reducing the lag time, overcoming the delayed crystallization observed in the unseeded ternary system.

**FTIR characterization of nucleobase assemblies**

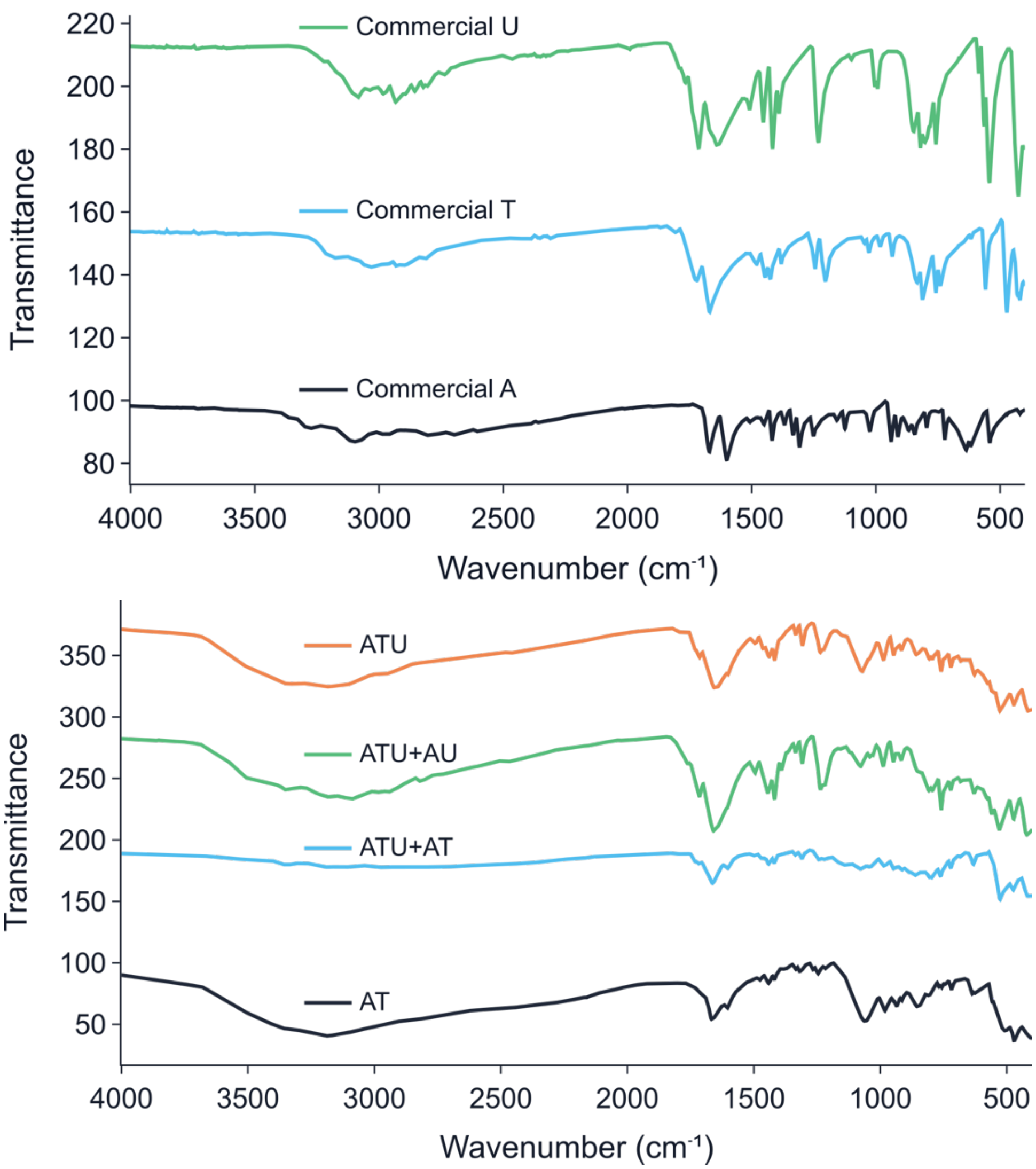

**Figure S11. FTIR characterization of nucleobase assemblies.** ATR-FTIR spectra of commercial individual nucleobases (top) and assembled binary and ternary samples (bottom). The FTIR spectra of all crystals showed similar peak positions, with no clear specific differences detected by this method.

**Energy-dispersive X-ray spectroscopy**

Elemental analysis of the A+T sample crystalline material was performed by energy-dispersive X-ray spectroscopy (EDX) using an extra-high-resolution scanning electron microscope equipped with an Oxford XMAX silicon drift detector. Measurements were performed at an accelerating voltage of 5 kV and a beam current of 6.2 nA. Spectra were processed using normalized analysis of all detected elements, with no peaks omitted. The spectrum was acquired for 30 s.

The EDX spectrum of the A+T crystalline sample showed the presence of carbon, nitrogen, oxygen, sodium, and phosphorus. The dominant contributions were oxygen and carbon, together with significant sodium and phosphorus signals, consistent with residual inorganic buffer/salt components associated with the dried crystalline sample.

**Table S5. Elemental analysis of dried A+T sample.**

| Element | X-ray line | Weight % | Weight % error | Atomic % |
|---|---|---|---|---|
| C | K | 23.93 | 0.51 | 32.79 |
| N | K | 5.32 | 0.67 | 6.26 |
| O | K | 40.64 | 0.48 | 41.82 |
| Na | K | 16.94 | 0.22 | 12.13 |
| P | K | 13.17 | 0.19 | 7.00 |
| **Total** | **100.00** | | | |

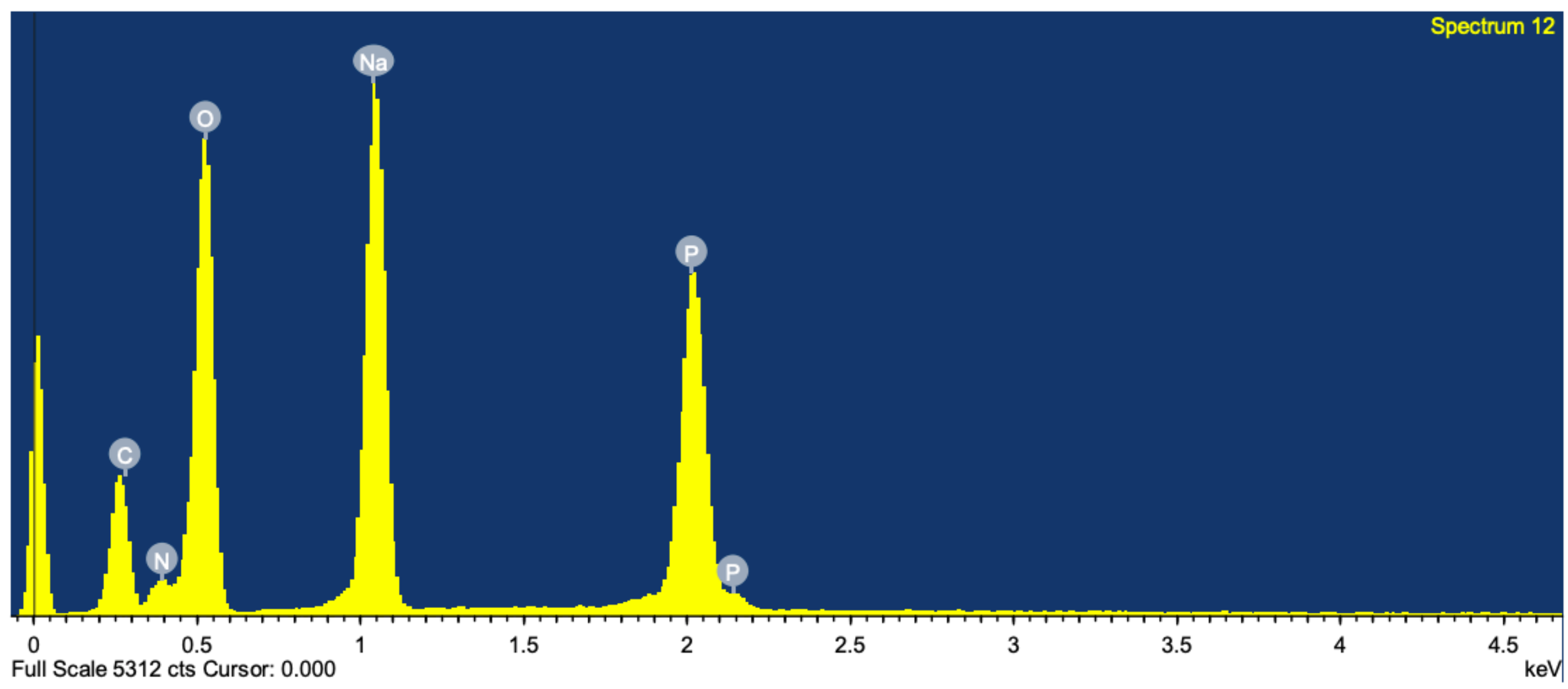

**Figure S12. EDX spectrum of the A+T crystalline sample.** Representative energy-dispersive X-ray spectroscopy (EDX) spectrum acquired from the dried A+T crystalline sample, showing signals assigned to C, N, O, Na, and P. The presence of Na and P indicates residual phosphate-buffer salts associated with the dried material. Quantitative elemental analysis was obtained from the normalized spectrum and is summarized in Table S5.

## HPLC Quantification

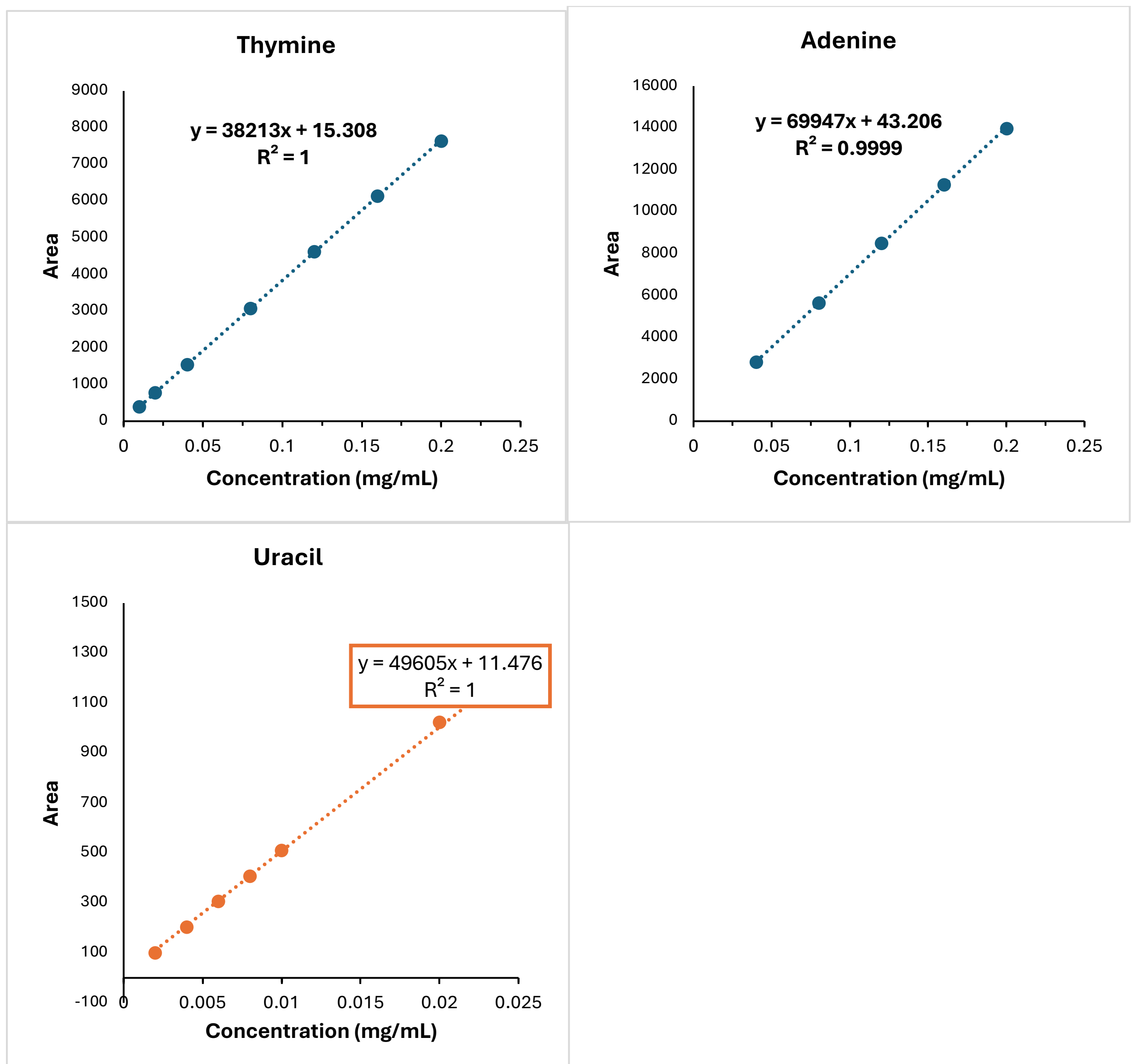

**Figure S13. Calibration curves for HPLC analysis taken from standard solutions of thymine, adenine, and uracil.**

**Table S6. Quantification of nucleobase composition in redissolved crystalline fractions.** Concentrations of A, T, and U were determined from two independent samples ('a' and 'b'), each measured at two injection concentrations (0.2 and 0.1 mg/mL total solid phase), as indicated. Values are reported as calculated concentrations in mM, with the average and standard deviation shown for each compound.

**C (mM)**

| Sample | *Compound* | *a 0.1* | *a 0.2* | *b 0.1* | *b 0.2* | *Ave* | *SD* |
|---|---|---|---|---|---|---|---|
| A+T | A | 0.402 | 0.414 | 0.457 | 0.460 | 0.433 | 0.030 |
| | U | NA | NA | NA | NA | NA | NA |
| | T | 0.208 | 0.214 | 0.236 | 0.238 | 0.224 | 0.015 |
| A+T+U | A | 0.382 | 0.390 | 0.340 | 0.394 | 0.376 | 0.025 |
| | U | 0.167 | 0.142 | 0.149 | 0.152 | 0.153 | 0.011 |
| | T | 0.279 | 0.286 | 0.235 | 0.250 | 0.262 | 0.024 |
| A+T+U+ A+T seeds | A | 1.208 | 1.218 | 0.643 | 0.644 | 0.928 | 0.329 |
| | U | 0.040 | 0.037 | 0.022 | 0.021 | 0.030 | 0.010 |
| | T | 0.596 | 0.605 | 0.316 | 0.318 | 0.459 | 0.164 |